\documentclass[twocolumn,           
               showpacs,            
               preprintnumbers,     
               aps,                 
               prd,                 
               a4paper,             
               groupedaddress,      
               nofootinbib,         
               tightenlines,        
               floats,floatfix      
               ]{revtex4}

\usepackage{graphicx}
\usepackage{latexsym}
\usepackage{amsmath,amssymb}        
\usepackage[draft=false]{hyperref}

\begin{document}


\title{Accretion disk onto boson stars: a way to supplant black holes 
candidates}

\author{F. Siddhartha Guzm\'an}

\affiliation{Instituto de F\'{\i}sica y Matem\'{a}ticas, Universidad 
Michoacana de San Nicol\'as de Hidalgo. Edificio C-3, Cd. Universitaria. 
C. P. 58040 Morelia, Michoac\'{a}n, M\'{e}xico.}



\begin{abstract}
The emission spectrum from a simple accretion disk model around a compact 
object is compared for the cases of a black hole (BH) and a boson star 
(BS) playing the role of the central object. It was found in the past that 
such a spectrum presents a hardening at high frequencies; however, here it 
is shown that the self-interaction and compactness of the BS have 
the effect of softening the spectrum, the less compact the star is, the 
softer the emission spectrum at high frequencies. Because the mass of the 
boson fixes the mass of the star and the self-interaction the compactness 
of the star, we find that, for certain values of the BS parameters, it is 
possible to produce similar spectra to those generated when the central 
object is a BH. This result presents two important implications: (i) using 
this simple accretion model, a BS can supplant a BH in the role of compact 
object accreting matter, and (ii) within the assumptions of the present 
accretion disk model we do not find a prediction that could help 
distinguish a BH from a BS with appropriate parameters of mass and 
self-interaction.
\end{abstract}

\maketitle


Since boson stars (BSs) were found to be stationary solutions of the 
Einstein-Klein-Gordon (EKG) system of equations \cite{ruffini,kaup} there 
have been two approaches for their study in astrophysical grounds: i) 
proof their existence through measurable predictions and ii) use BSs as 
toy models for problems requiring an exotic matter component. Within the 
first aim some topics have been developed: prediction of particular 
signatures in gravitational wave emission of perturbed boson stars 
\cite{qnm-japoneses,ryan,caltech,bondarescu}; prediction of differences in 
the emission spectrum in accretion disks around supermassive BSs vs that 
from BHs \cite{diego,diego-acc,rees}. The second line of 
research involves proposing gigantic BSs as galactic dark halos 
\cite{jae-koh-dm,schunck-dm,schunck-phd,guzmanlau2003}; BSs as 
alternative to MACHOS \cite{machos-schunck} and BSs as solutions of 
tensor-scalar theories \cite{van-der,diego-jbd,bala-phd}. 
Around these two fronts the understanding of BSs has been widely 
developed, and the most important results can be found in two recent 
reviews by Jetzer \cite{jetzer} and Schunck and Mielke \cite{cqg-review}. 
In this manuscript we focus on the first approach, and try to provide a 
picture of the properties that make a BS look like a BH, which we 
consider to be important because observations are at the verge of proving 
-or not- that black hole candidates (BHCs) have a horizon or not and when 
such debate involves BSs \cite{rees,abramowics,narayan}. Firstly we 
present the fundamentals on Boson Stars as follows.

Boson stars are stationary solutions of Einstein's equations $G_{\mu \nu} 
= 8\pi G T_{\mu \nu}$, where $T_{\mu \nu} = \frac{1}{2}[\partial_{\mu} 
\phi^{*} \partial_{\nu}\phi + \partial_{\mu} \phi \partial_{\nu}\phi^{*}] 
-\frac{1}{2}g_{\mu \nu} [\phi^{*,\alpha} \phi_{,\alpha} + V(|\phi|^2)]$ 
is the stress-energy tensor of a complex scalar field $\phi$, 
$g_{\mu\nu}$ is the metric of the space-time and $V$ is the 
potential of the field, which for the present case is restricted to have 
the form $V(|\phi|) = \frac{1}{2}m^2 |\phi|^2 + \frac{\lambda}{4}|\phi|^4$ 
where $m$ is the mass of the field and $\lambda$ its self-interaction.
Whether or not the scalar field plays the role of a fundamental 
spinless particle, one can always consider it appears in an effective 
theory with lagrangian density ${\cal L} =  -\frac{R}{16 \pi G} + g^{\mu 
\nu}\partial_{\mu} \phi^{*}\partial_{\nu}\phi + V(|\phi|)$. This mean 
field approach helps at avoiding the formulation of quantum field theory 
in a curved background.

Although axi-symmetric BS solutions have been found, we restrict here 
to the spherically symmetric case and choose the line element of 
space-time in Schwarzschild coordinates to be $ds^2=-\alpha(r)^2dt^2 + 
a(r)^2dr^2 + r^2 d\Omega^2$. Assuming the field to have a time 
dependence $\phi(r,t) = \phi_0(r) e^{-i \omega t}$ the stress energy tensor 
is time independent, which implies that the space-time is stationary and 
therefore that the metric functions $\alpha$ and $a$ depend only on the 
radial coordinate $r$. Under this assumptions the Einstein-Klein-Gordon 
equations are:

\begin{eqnarray}
\frac{a'}{a} &=& \frac{1-a^2}{2r}+\frac{r}{2}
\left[\phi_{0}^{2}\frac{a^2}{\alpha^2}+\phi_0'{}^{2} + a^2 (\phi_{0}^{2} 
+ \frac{1}{2}\Lambda \phi_{0}^{4})
\right]\label{sphericalekga-sc}\\
\frac{\alpha'}{\alpha} &=& \frac{a^2-1}{r} + \frac{a'}{a} -
ra^2\phi_{0}^{2}(1+\frac{1}{2}\Lambda\phi_{0}^{2})\label{sphericalekgb-sc}\\
\phi_0 '' &+& \phi_0 ' \left( \frac{2}{r} + \frac{\alpha '}{\alpha} -
\frac{a'}{a}\right) + \phi_0 \frac{a^2}{\alpha^2} -
a^2 (1 + \Lambda \phi_{0}^{2}) \phi_0 =0\label{sphericalekgc-sc}
\end{eqnarray}

\noindent where we have used the rescaled variables
$\phi_0 \rightarrow \sqrt{4\pi G} \phi_0$, $r \rightarrow mr$, 
$t \rightarrow \omega t$, $\alpha \rightarrow \frac{m}{\omega}\alpha$ and 
$\Lambda = \frac{\lambda}{4 \pi G m^2}$; primes 
indicate derivative with respect to the radial coordinate.
The system (1-3) is a set of coupled ordinary differential equations to be
solved under the conditions $a(0)=1$, $\phi_0(0)$ finite and
$\phi_0\prime(0)=0$ in order to guarantee regularity and spatial flatness
at the origin, and $\phi_0(\infty)=\phi_0 \prime(\infty)=0$ in order to
ensure asymptotic flatness at infinity as described in
\cite{ruffini,seidel90,seidel98,gleiser,scott}; these conditions reduce 
the system (1-3) to an eigenvalue problem for $\omega$. The solution was 
calculated numerically using finite differencing and a shooting routine 
that searched $\omega$.


\begin{figure}[htp]
\includegraphics[width=8cm]{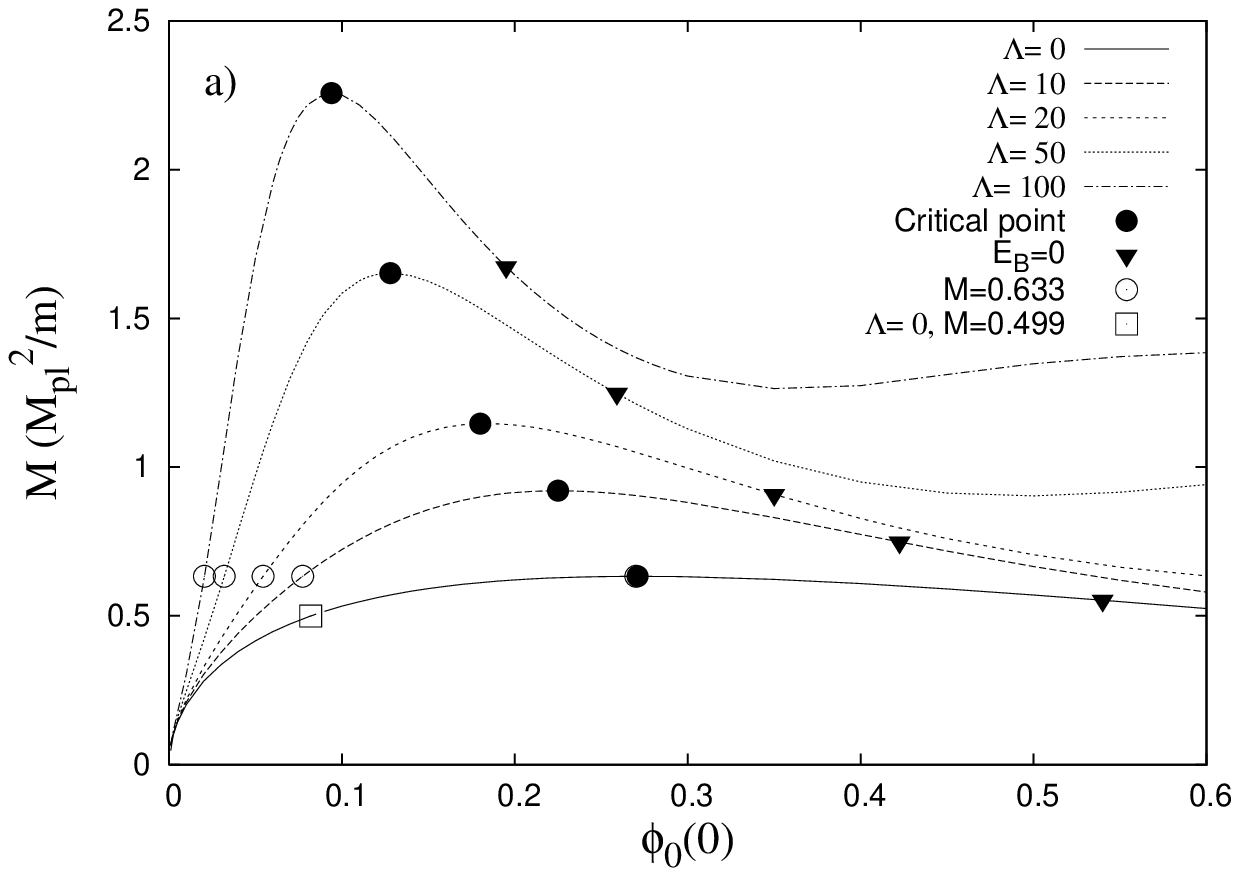}
\includegraphics[width=8cm]{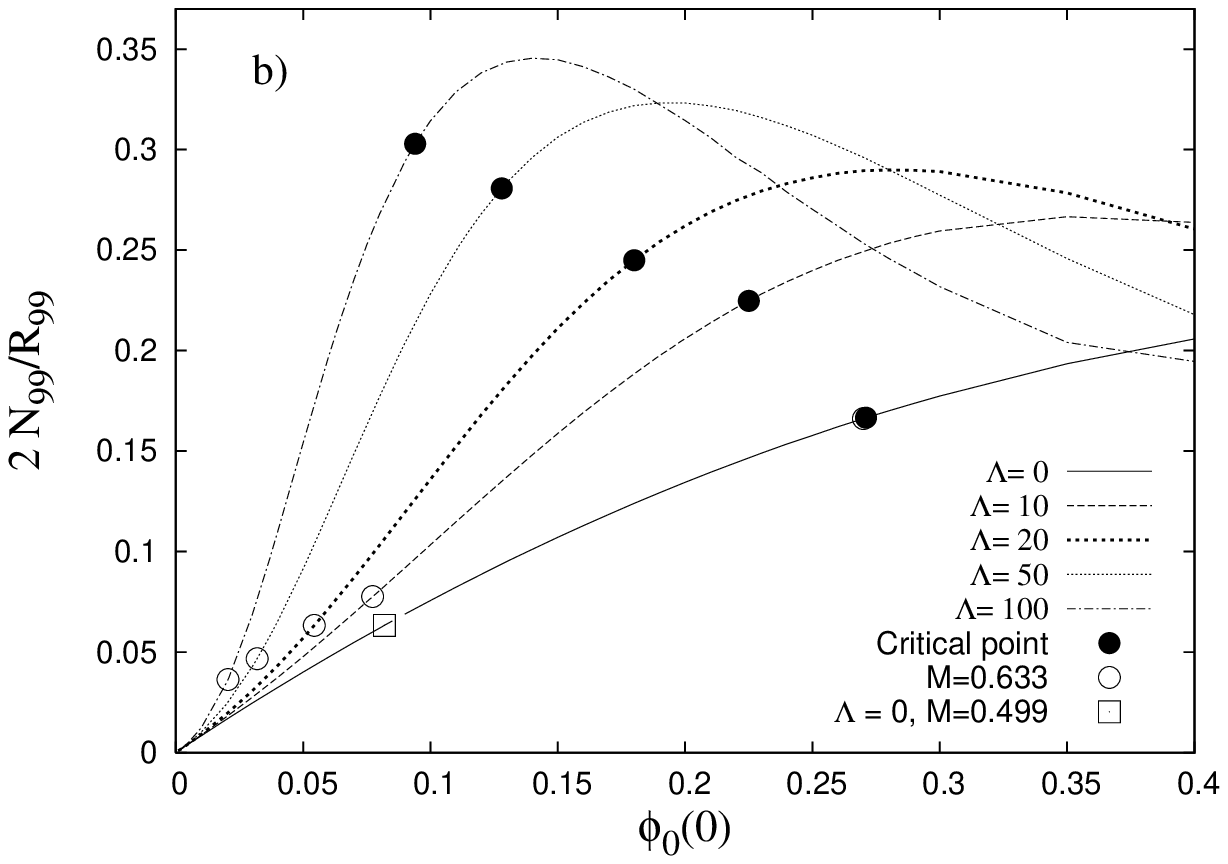}
\caption{\label{fig:equilibrium} (a) Sequences of equilibrium 
configurations for different values of $\Lambda$ are shown as a function 
of the central value of the scalar field $\phi_0(0)$. The circles 
indicate the critical point that divides the stable from the unstable 
solutions. The inverted triangles indicate the point at which the binding 
energy is zero. Those configurations between the circles and the 
triangles (along each sequence) collapse into black holes even under 
infinitesimal perturbations (see \cite{guzman2004} for the full tracking 
of the formation of an event horizon out of an unstable boson star). 
Configurations to the right of the triangles disperse away. (b) The 
compactness of each solution is shown. The critical point is also marked, 
since the interesting configurations would be those boson stars that are 
stable (to the left of the circles). In both plots, the transparent 
circles indicate the special set of configurations we calculate the 
emission spectrum for in what follows; in the case of $\Lambda=0$ such 
configuration is very close to the critical point and it is difficult to 
look at, but it is in there. The configuration marked with a square 
corresponds to a configuration with $\Lambda=0$, $M=0.499$ and 
compactness 0.06332; this compactness is the same as that of 
configuration with $\Lambda=20$ and $M=0.633$, see text below to take 
advantage of this remark.} 
\end{figure}

The solutions to this set of equations define sequences of 
equilibrium configurations like those shown in Fig. 1(a). In the curves 
two important points for each value of $\Lambda$ are marked: i) the 
critical point -marked with a filled circle- indicating 
the threshold between the stable and unstable branches of each sequence, 
that is, configurations to the left of this point are stable and those to 
the right are unstable as found through the analysis of perturbations 
\cite{gleiser,scott}, catastrophe theory \cite{catastrophe} and full 
non-linear evolution of the evolution equations \cite{seidel90,seidel98} 
and ii) the point at which the binding energy $E_B 
= M-Nm = 0$ marked with an inverted filled triangle, where 
$N=\int j^0 d^3 x = \int 
\frac{i}{2}\sqrt{-g}g^{\mu\nu}[\phi^{*}\partial_{\nu}\phi - \phi 
\partial_{\nu}\phi^{*}]d^3 x$ is the number of particles and $M = 
(1-1/a^2)r/2$ evaluated at the outermost point of the numerical domain is 
the Schwarzschild mass; the configurations between the instability 
threshold and the zero binding energy point collapse into 
black holes whereas those to the right disperse away as shown in 
\cite{guzman2004}. The units for the $M$ and $N$ are given in 
$M_{pl}^{2}/m$, where $M_{pl}$ is the Planck mass and $m$ is the mass 
of the boson.

Since the mass of the configurations in Fig. 1(a) scales with $m$, 
the original use of the self-interaction $\Lambda$ was to allow bigger 
masses even if the mass parameter $m$ was fixed \cite{colpi} and thus BS 
configurations seemed to be -gravitationally- similar to compact objects 
like neutron stars \cite{colpi}. In the present approach neither $m$ nor 
$\Lambda$ will be considered related to any type of possibly existing 
particle, instead they are thought of as free parameters that permit 
faking compact objects. We prefer to focus on another property of BSs: the 
compactness. In Fig. 1(b) the compactness of equilibrium configurations is 
shown. Provided BSs have no defined surface we consider that the radius 
containing 99\% ($R_{99}$) of the total particle number (following 
\cite{seidel90,seidel98} where 95\% was considered instead) is a 
reasonable place where to measure the gravitational field of the star; 
thus the compactness plotted in Fig. 1(b) is defined as $2 N_{99}/R_{99}$, 
where $N_{99}$ is the number of particles integrated up to $R_{99}$. 
One could feel tempted to believe that the more compact a BS is 
more similar to a BH seems, however as shown below this happens only in 
terms of how deep the trajectories of test particles can be.\\

{\it Time-like geodesics.} Given the line element $ds^2 = -\alpha(r)^2 
dt^2 + a(r)^2dr^2 + r^2d\Omega^2$, the equation for time-like geodesics 
followed by test particles on the equatorial plane reads:

\begin{equation}
\dot{r}^2 + \frac{1}{a^2}\left(1 + \frac{L^2}{r^2}\right) = 
\frac{E^2}{\alpha^2 a^2},
\label{eq:geodesics}
\end{equation}

\noindent  where $L^2 = r^4\dot{\varphi}^2$ and $E^2 = - 
\alpha^2 \dot{t}^2$ are the squared angular momentum and energy 
at spatial infinity, which are the conserved quantities of the test 
particle related to the independence on the azimuthal angle $\varphi$ and 
$t$ of the space-time respectively; an overdot indicates derivative with 
respect to the proper time of the test particle. The geodesics for a 
Schwarzschild BH are given by (\ref{eq:geodesics}) with the values 
$\alpha^2=a^{-2}=(1-\frac{2M}{r})$. Since equation 
(\ref{sphericalekgb-sc}) for $\alpha$ is linear, we rescale this function 
so that at infinity $\alpha(r \rightarrow \infty) = 1/a(r \rightarrow 
\infty)$, thus at infinity $\alpha^2 = a^{-2}$, which implies the 
coefficient of $E^2$ in (\ref{eq:geodesics}) is one for BSs, which also 
happens for the BH.

\begin{figure}[htp]
\includegraphics[width=8cm]{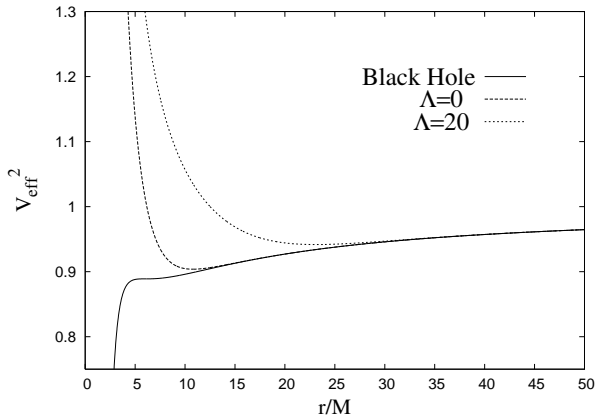}
\caption{\label{fig:effective-p} Here it is shown the effective potential 
that a particle with $L^2 = 12 M^2$ ``feels'' for two configurations of 
the same mass $M=0.633$ and $\Lambda=0,20$, and for a black hole of the 
same mass. At $r=6M$ it is possible to see the ISCO for a test particle on 
a BH 
space-time. The interesting result is that for the more compact stars 
test particles can travel deeper inside the star. What is possible 
to infer is that freely falling observers cannot distinguish whether there 
is a BH or a BSs from $r=30M$ on.} 
\end{figure}

At this point it is convenient to draw the effective potential 
$V_{eff}^{2} = \frac{1}{a^2}\left(1 + \frac{L^2}{r^2}\right)$ for a given 
angular momentum of the test particle. In Fig.~\ref{fig:effective-p} we 
show $V_{eff}^{2}$ for different values of $L$ and compare with the 
potentials due to a BH. It happens that the effective potentials for a BS 
depend on the compactness of the star only and are independent of the 
self-interaction; instead of showing this fact we present the potentials 
for configurations with the same mass ($M=0.633 (M^{2}_{pl}/m)$) and 
$\Lambda =$ 0, 20 with compactness 0.1666 and 0.06332 
respectively. The location of the typical potential barrier for test 
particles on a BS space-time \cite{diego,caltech,guzman2005} changes from 
one case to the other: the less compact the star ($\Lambda=20$) the 
farer the location of the barrier from the center.

\begin{figure}[htp]
\includegraphics[width=8cm]{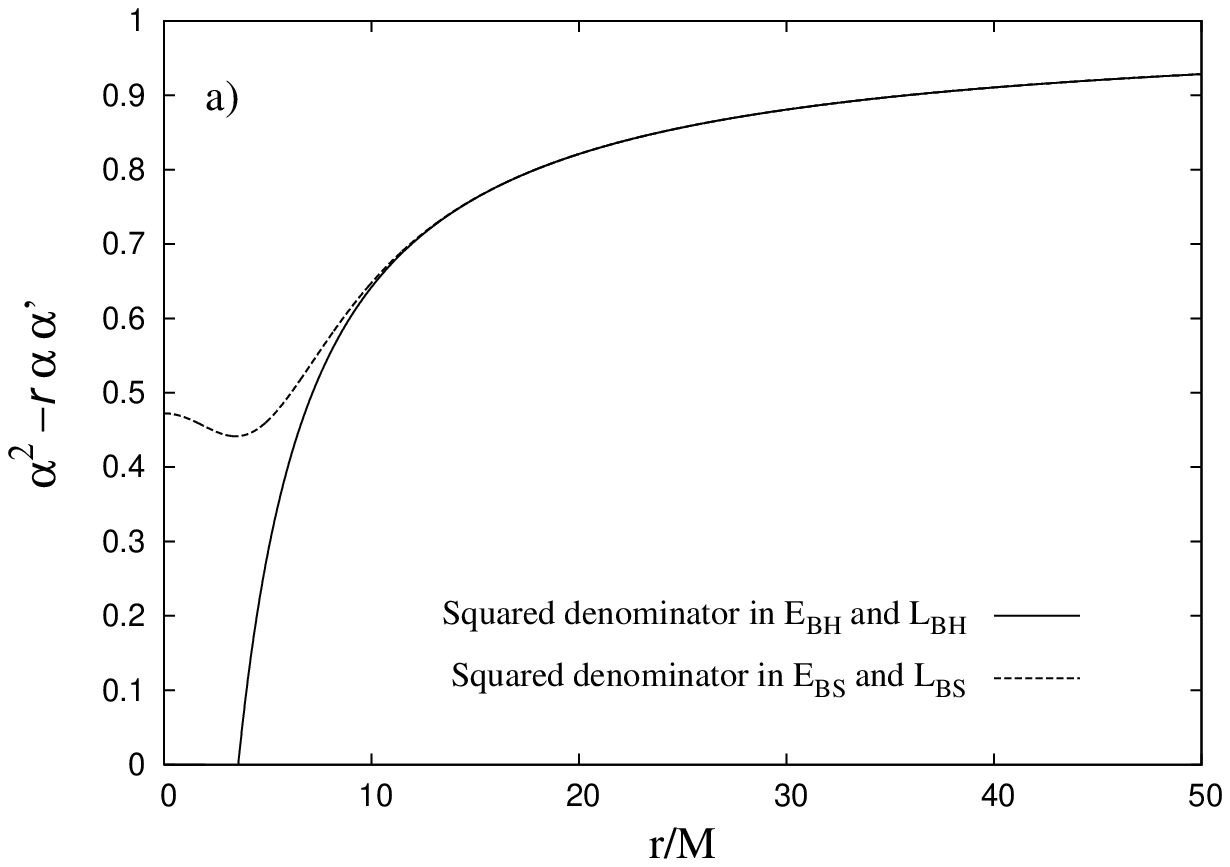}
\includegraphics[width=8cm]{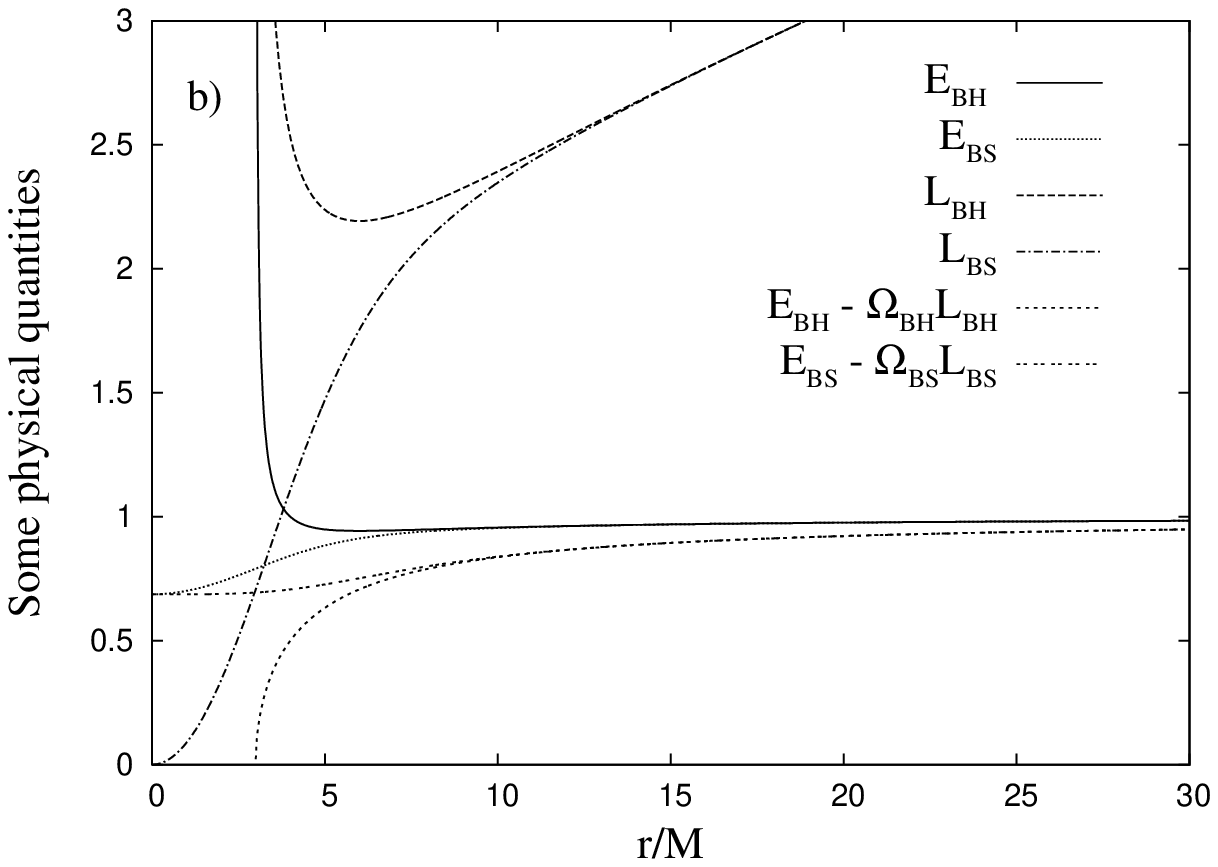}
\caption{\label{fig:EyL} a) The squared denominator of $E$ and $L$ in 
(\ref{eq:EyL}); the star chosen is the one with $\Lambda=0$, central field 
$\phi(0)=0.271$ and mass $M=0.633 (M_{pl}^{2}/m)$. b) Physical 
quantities used for the calculation of $D(r)$ (see below); the solid line 
indicates these quantities for a black hole of the same mass as the star. 
The results in this plot coincide with those obtained in 
\cite{diego-acc}.} 
\end{figure}

It is known that BSs admit stable circular orbits for every radius 
\cite{diego-acc}. Thus, here we do not touch that problem and restrict to 
calculate $E$ and $L$ in terms of the metric functions. In order to do so 
we rewrite (\ref{eq:geodesics}) for $\dot{r}=0$ and demand $dV/dr=0$ with 
$V=\left[ \left( 1 + L^2/r^2 \right) - E^2/\alpha^2\right]/a^2$. This 
condition implies 

\begin{equation}
E = \frac{\alpha^2}{\sqrt{\alpha^2 - r\alpha\alpha^{\prime}}}, ~~~\&~~~
L = \sqrt{\frac{r^3\alpha\alpha^{\prime}}{\alpha^2 - 
r\alpha\alpha^{\prime}}}
\label{eq:EyL}
\end{equation}

\noindent where $\alpha^2 - r\alpha\alpha^{\prime} > 0$ all the way as 
shown in Fig.~\ref{fig:EyL}(a) for a typical BS. 
Before proceeding to the calculation of the emission spectrum we need the 
angular velocity of a test particle given by $\Omega = \sqrt{\frac{\alpha 
\alpha^{\prime}}{r}}$, which is a regular function for all values of $r$. 
In Fig~\ref{fig:EyL}(b) the values of $E$, $L$ and the useful $E-\Omega L$ 
are shown for the $M=0.633$, $\Lambda=20$ BS and a BH of the same mass.


The accretion disk model is that of a geometrically thin, optically thick, 
steady accretion disk. The power per unit area generated by such a disk 
rotating around a central object is given by \cite{page-thorne,diego-acc}:

\begin{equation}
D(r) = \frac{\dot{M}}{4\pi r}\frac{\alpha}{a}\left(-\frac{d\Omega}{dr}
\right)
\frac{1}{(E-\Omega L)^2} \int^{r}_{r_{i}}(E-\Omega L)\frac{dL}{dr}dr 
\label{eq:power}
\end{equation}

\noindent where $\dot{M}$ is the accretion mass rate and $r_{i}$ is the 
inner edge of the disk, keeping in mind that $r$ is in units of $1/m$. For 
black holes this radius is assumed to be at the ISCO ($r=6M$) of the hole. 
Since BSs allow circular orbits in the whole spatial domain we consider 
that $r_{i} = 0$ for BSs. Furthermore, assuming it is possible to define a 
local temperature we use the Steffan-Boltzmann law so that $D(r) = \sigma 
T^4$, where $\sigma=5.67 \times 10^{-5}~ erg~s^{-1}~cm^{-2}~K^{-4}$ is the 
Steffan-Boltzmann constant. Now, considering the disk emits as a black 
body, we use the dependence of $T$ on the radial coordinate 
and therefore the luminosity $L(\nu)$ of the disk and the flux $F(\nu)$ 
through the expression for the black body spectral distribution:

\begin{equation}
L(\nu) = 4\pi d^2 F(\nu) =
\frac{16 \pi h}{c^2} \cos (i) \nu^3 \int^{r_f}_{r_i} 
\frac{rdr}{e^{h\nu/kT} - 1}
\label{eq:luminosity}
\end{equation}

\noindent where $d$ is the distance to the source, $r_i$ and $r_f$ 
indicate the location of the inner and outer edges of the disk, $h=6.6256 
\times 10^{-27}~ erg~s$ is the Planck constant and $k = 1.3805 \times 
10^{-16}~ erg~K^{-1}$ is the Boltzmann constant and $i$ is the 
disk inclination. The algorithm to construct the emission spectrum for 
such a model of accretion disk around a BS and a BH is as follows: \\

\noindent 1) Define the space-time functions $a$ and $\alpha$ by choosing 
one of the equilibrium configurations in Fig.~\ref{fig:equilibrium} and 
calculate $M$.\\
2) Define the metric of the equivalent BH through $\alpha_{BH} = 
\sqrt{1-2M/r}$ and $a_{BH} = 1/\sqrt{1-2M/r}$.\\
3) Calculate the angular velocity, angular momentum and energy of a test 
particle for both space-times $\Omega_{BS, BH}$, $L_{BS,BH}$, 
$E_{BS,BH}$.\\
4) Use such quantities to calculate the power emitted in both cases 
$D_{BS}(r)$ and $D_{BH}(r)$ defined in (\ref{eq:power}).\\
5) Calculate the temperature of the disk in both cases 
$T_{BS}(r) = (D_{BS}(r)/\sigma)^{1/4}$ and 
$T_{BH}(r) = (D_{BH}(r)/\sigma)^{1/4}$.\\
6) Use such temperature to integrate the luminosity (\ref{eq:luminosity})
$L_{BS}(\nu)$ and $L_{BH}(\nu)$ for several values of $\nu$.\\

\begin{figure}[htp]
\includegraphics[width=8cm]{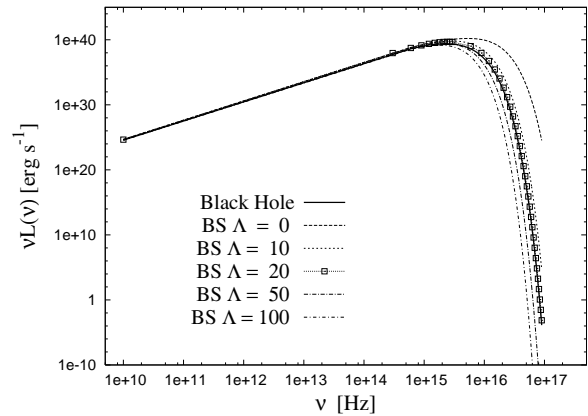}
\caption{\label{fig:spectrum} The emission spectrum from a central star 
with mass $M=0.633 (M_{pl}^{2}/m)$ and different values of $\Lambda$. The 
solid line is obtained when a black hole is considered as the central 
object. The curve for $\Lambda=0$ coincides with the one obtained in \cite{diego-acc} 
and shows the hardening of the spectrum at high frequencies. 
The case $\Lambda=20$ is particularly interesting because it seems to 
match the spectrum obtained with the BH.} 
\end{figure}

We choose a configuration with $M=0.633 (M_{pl}^2/m)$ because it has been 
used in the past (e.g. \cite{diego-acc}), but it could be any other; the 
usual reason to use this configuration is that for $\Lambda=0$ it 
corresponds to the most compact configuration in the stable branch (see 
Fig.~\ref{fig:equilibrium}). We take the physical parameters used in 
\cite{diego-acc}, where $m=1.2 \times 10^{-25}$GeV, $\dot{M}=0.633 
(M_{pl}^2/m) = 2.8 \times 10^{6}M_{\odot}$, $i=60^{0}$, which we use 
here too. We used $r_f$ = 30 and 50$M$ and found no qualitative difference 
in the spectrum. As a test in the present calculations we show here the 
results found in \cite{diego-acc} for such system in 
Fig.~\ref{fig:spectrum} as a particular case, which is interpreted as a 
possible signature of BSs because there is a hardening of the spectrum at 
high frequencies.

Nevertheless, what we do here is to repeat our algorithm for the same mass 
of the object but different values of $\Lambda$ keeping the same value of 
the mass (the configurations marked with simple circles in Fig. 1). The 
result can be found in Fig.~\ref{fig:spectrum}, where the spectrum is 
softened when increasing $\Lambda$, in fact we notice that for 
$\Lambda=20$ the spectrum could perfectly be that of the equivalent black 
hole; we dub this particular configuration the Star $Y$ and remark 
its compactness $2M_{99}/R_{99}=0.06332$. Under the present 
model of accretion disk it would be rather difficult to distinguish 
between a black hole and a boson star with mass $M= 2.8 \times 10^{6} 
M_{\odot}$ under the parameters $m=1.2 \times 10^{-25}$GeV and 
$\Lambda=20$ compared with the case $\Lambda=0$ where $\nu L(\nu)$ is 
orders of magnitude bigger for the BS than for the BH around the 
$10^{16}Hz$ window.\\

\begin{figure}[htp]
\includegraphics[width=8cm]{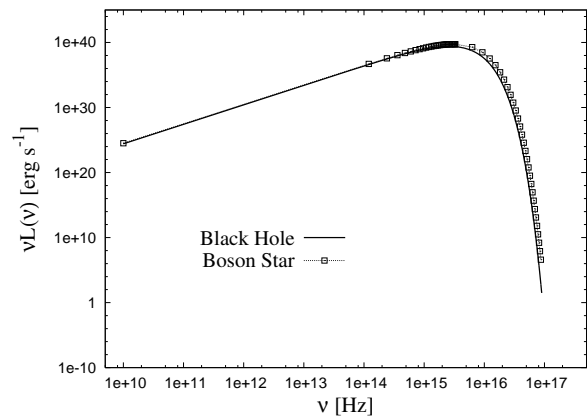}
\caption{\label{fig:spectrum-bis} The emission spectrum from a disk with a 
central star with mass $M=0.499 (M_{pl}^{2}/m)$ and $\Lambda=0$. In order 
to keep $M$ equal to that of configurations in Fig.~\ref{fig:spectrum} it 
is necessary to set $m=0.499/0.633 \times 1.2 \times 10^{-25}$GeV, 
otherwise we could construct a correct spectrum with the incorrect 
gravitational mass. The similarity between the spectrum due to the BS and 
to the BH is not as impressive as that of the Star $Y$, but the hardening 
of the spectrum is not as strong as that of $\Lambda=0$ in the previous 
plot. We simply used a star with no self-interaction and compactness 
$2N_{99}/R_{99}=0.06332$ as that of Star $Y$.} 
\end{figure}

Once the algorithm is settled down we try one last experiment: we choose a 
$\Lambda=0$ star with the same compactness as that of star $Y$, such 
configuration appears in Fig. 1 marked with a square. The mass of such 
star is $M=0.499 (M_{pl}^{2}/m)$; in order to achieve a mass equal to that 
of configurations of Fig.~\ref{fig:spectrum} we need to use 
$m=0.499/0.633 \times 1.2 \times 10^{-25}$GeV. The resulting spectrum is 
shown in Fig.~\ref{fig:spectrum-bis} where once again it is possible to 
nearly match the spectrum corresponding to that of a BH with the same 
mass, but this time without the need of self-interaction.\\

{\it Discussion and conclusions.} There are two main results: 1) BSs can 
play the role of BHCs and 2) the emission spectrum is not helpful at 
determining whether or not the BHC is a BS or a BH and in 
consequence it is not enough to say whether the compact object has a 
horizon or not within the accretion model of the present approach. There 
are two ways of faking the spectrum of the disk around a BH: i) by using 
an appropriate value of $\Lambda$ for a given $m$ and ii) by using an 
appropriate value of $m$ with $\Lambda=0$.\\

We expect the results presented here are disk model dependent, however 
the study of other different accretion disk models could show a similar 
behavior. Nevertheless, at this point one can conclude that the spectrum 
is not enough to determine whether a compact object is a BS or not, and in 
any case $m$ and $\Lambda$ remain free parameters that could help at 
matching an emission spectrum.\\

For instance, in \cite{rees} the accretion of ordinary nucleonic gas on 
black hole candidates is analyzed by assuming that the BHC is either a 
perfect fluid star or a boson star instead of a black hole. The result 
indicates that for such cases Type I X-ray bursts should be observed 
because the accreting gas particles pile up on the surface of the stars, 
since in both type of stars the Schwarzschild radius lies inside the star. 
Because such bursts are not observed, these two types of stars are ruled 
out as BHCs and it is considered that the BHC is a black hole. 
Nonetheless, the assumption that BSs have surface is not correct and a 
test particle would not be subject to mechanical forces due to the 
presence of a surface; thus the analysis in \cite{rees} should be redone 
under such consideration.\\

Finally, it could be that under different accretion disk models, 
particular observable BS signatures could appear in the electromagnetic 
spectrum. At the same time, it is remarkable the effort invested in the 
prediction of BS signatures through gravitational wave signals due to 
particles surrounding a BS \cite{caltech} and due to full 3D perturbations 
\cite{bondarescu}.


\acknowledgments
The author thanks Diego F. Torres and L. Arturo Ure\~na-L\'opez for 
criticism, comments and ideas. This research is partly granted by 
CIC-UMSNH-4.9 and PROMEP-UMICH-PTC-121. The runs were carried out in the 
Ek-bek cluster of the ``Laboratorio de Superc\'omputo Astrof\'{\i}sico 
(LASUMA)'' at CINVESTAV-IPN. The code was developed in hardware from the 
Center for Computation and Technology at Louisiana State University.


\end{document}